\documentclass[12pt,preprint]{aastex}
\usepackage{emulateapj5}
\usepackage{epsfig}

\newcommand{\mincir}{\raise
-2.truept\hbox{\rlap{\hbox{$\sim$}}\raise5.truept\hbox{$<$}\ }}
\newcommand{\magcir}{\raise
-2.truept\hbox{\rlap{\hbox{$\sim$}}\raise5.truept\hbox{$>$}\ }}

\newenvironment{inlinefigure}{%
\def\@captype{inlinesfigure}%
\noindent\begin{minipage}{\linewidth}\begin{center}}
{\end{center}\end{minipage}\smallskip}
\makeatother

\begin{document}

\title{Luminous X-ray AGN in Clusters of Galaxies}
\author{E. Koulouridis\altaffilmark{1}, M. Plionis\altaffilmark{1,2}}

\altaffiltext{1}{Institute of Astronomy \& Astrophysics, National 
Observatory of Athens, I.Metaxa \& B.Pavlou, P.Penteli 152 36, 
Athens, Greece, e-mail:ekoulour@astro.noa.gr} 

\altaffiltext{2}{Instituto Nacional de Astrof\'{\i}sica \'Optica y 
Electr\'onica, 72840, Puebla, Pue, Mexico}

\begin{abstract}
We present a study of X-ray AGN overdensities in 16 Abell clusters, within the redshift range
$0.073<z<0.279$, in order to investigate the effect of the hot inter-cluster 
environment on the triggering of the AGN phenomenon.
The X-ray AGN overdensities, with respect to the field expectations,
were estimated for sources with $L_x\ge 10^{42}$ erg s$^{-1}$ 
(at the redshift of the clusters) 
and within an area of 1 $h^{-1}_{72}$ Mpc radius (excluding the core).
To investigate the presence or not of a true enhancement of luminous X-ray AGN
in the cluster area, we also derived the corresponding optical galaxy overdensities,
using a suitable range of $r$-band magnitudes.
We always find the latter to be significantly higher (and only in two cases
roughly equal) with respect to the corresponding X-ray
overdensities. Over the whole cluster sample, the mean X-ray
point-source overdensity is a factor of $\sim 4$ less than that
corresponding to bright optical galaxies, a difference which is 
significant at a $>0.995$ level, as indicated by an appropriate $t$-student test.
We conclude that the triggering of luminous X-ray AGN in rich clusters is
strongly suppressed. 
Furthermore, searching for optical  {\em Sloan Digital Sky Survey} (SDSS) 
counterparts of all the X-ray sources, 
associated with our clusters, we found that about half
appear to be background QSOs, while others are 
background and foreground AGN or stars.
The true overdensity of X-ray point sources, associated to the clusters, is
therefore even smaller than what our statistical approach revealed.
\end{abstract}

\keywords{galaxies: active, galaxies: clusters: general, X-rays: galaxies, X-rays: galaxies: clusters, X-rays: general}

\section{Introduction}
There is a growing body of studies investigating the effect of
the environment on the nuclear activity of galaxies and on the possible
triggering mechanisms of the AGN
phenomenon (e.g., Dultzin-Hacyan et al. 1999; Koulouridis et al. 2006;
2009; Sorrentino et al. 2006; Gonz\'alez et al. 2008; Choi et
al. 2009; Silverman et al. 2009; Lee et al. 2009; von der Linden et
al. 2009; Padilla, Lambas \& Gonz\'alez 2009
and references therein). 

One particular research direction is the study of X-ray AGN,
as a function of environment
, since undoubtedly, one of the best AGN identification methods is through X-ray
observations.
Clusters of galaxies offer an ideal target for this type of
study and indeed most X-ray based studies report overdensities of X-ray
sources in clusters, with respect to the field 
(e.g., Cappi et al. 2001; Molnar et al. 2002; 
D'Elia et al. 2004; Branchesi, et al. 2007;
Galametz et al. 2009; Gilmour et al. 2009). 

There are attempts to substantiate such results with spectroscopic
data and indeed various studies have verified the existence of a large
population of X-ray AGN in clusters of galaxies and its probable
evolution with redshift (eg. Martini et al. 2002; 
Johnson et al. 2003; Martini et al. 2007, 2009; van Breukelen et al. 2009). 
Some of the previous studies
have also shown that only a small fraction of the X-ray sources, 
associated with clusters of galaxies, possess optical
emission line ratios characteristic of AGN (see also
Davis et al. 2003, Finoguenov et al. 2004).
Similarly, starting from optical spectroscopy, 
Arnold et al. (2009) found that 
14 out of 144 galaxy members of 11 clusters and groups are
AGN, but with only one being detected in X-rays. 
A similar trend of disconnection between X-ray and
optically selected AGN, in eight poor groups of galaxies, was also 
found by Shen et al. (2007). 

We believe that there is a major question still not adequately answered,
which is: {\em Is there an
enhancement of AGN activity and/or of its X-ray manifestation
in clusters of galaxies with respect to what expected from the obvious
optical galaxy overdensity?}
We attempt to address this question by 
searching not only for the existence or not of a cluster X-ray AGN excess, with respect
to the field, since such could possibly be expected on the basis of the
known excess of optical galaxies in clusters, but rather by 
investigating whether the X-ray AGN overdensity shows a relative
enhancement or suppression with respect to the corresponding optical
galaxy overdensity.

\section{Sample selection \& Methodology}
The clusters of galaxies used in this study 
were selected according to the following criteria:

\noindent
(a) They belong to the list of Abell clusters (Abell et al. 1989), to
ensure a relatively large number of member galaxies;
(b) They have been observed by {\sc Xmm-Newton} with an
exposure time, in each of the three detectors, of more than 10 ksec. 
This is to ensure adequate
photon-counts and to reach a relatively low flux-limit;
(c) The diffuse X-ray emission of the cluster, in the center
of the field, is not as strong as to make impossible the detection of
point sources located more than 0.5 $h_{72}^{-1}$ Mpc from the center, at
worst;
(d) They lie in the area covered by the SDSS, in order
to facilitate the search of optical counterparts of the X-ray point sources.

When this study began in 2006, 16 galaxy clusters with redshift 0.073
$\leq z \leq $ 0.279 were found to meet the above conditions. 
Our main aim was to calculate 
the density of X-ray sources in the clusters region and compare it with
the corresponding density of optical galaxies. The cluster list, their redshift and various characteristics,
determined in the present study, are shown in Table 1.

The corresponding 16 XMM cluster fields were
analyzed using the Science Analysis Software (SAS)
of the {\sc Xmm-Newton}. For the final
extraction of sources we merged the images from all three detectors
(MOS1, MOS2 and pn) in the energy range 0.5-8 keV (at the observers frame)
and selected the
sources located within a radius of 1 $h_{72}^{-1}$ Mpc from the center of the
cluster and 5$\sigma$ above the background. 
We then extracted the flux of the detected sources, its corresponding X-ray 
luminosity at the redshift of the cluster assuming a power-law spectrum with $\Gamma=1.7$ 
and the sensitivity of the detectors in each pixel of our
fields. Since the sensitivity varies greatly across the field of view,
mostly due to vignetting,
we constructed the sensitivity map for each XMM field in order
to eventually calculate accurately the
theoretically expected number of sources in each region of interest. 

Due to the fact that the diffuse cluster X-ray emission 
can be strong enough to hide point sources, we have excluded 
from our analysis the
central region of each cluster. This was determined by
adjusting a King's profile (King et al. 1962) to the diffuse emission,
and deriving the cluster core radius, which is different for each cluster
and ranges from $\sim$ 20 to 170 $h^{-1}_{72}$ kpc. 
In order to efficiently remove the influence of the diffuse emission
of the central cluster area, we consistently used for all clusters an
inner radial cutoff of 3$ \times r_c$. 
We however remind the reader that
there are indications, from {\sc Chandra} data, of an increasing AGN population towards the 
centers of clusters (eg., Ruderman \& Ebeling 2005; Martini et al. 2007; Gilmour et al. 2009), 
which we cannot however efficiently probe with {\sc Xmm-Newton}.

We then counted all X-ray sources with luminosity $L_x\ge 10^{42}$
erg/s ($N_x$) located within the effective cluster
region (ie., the annulus between $3\times r_c$ and 1 $h^{-1}_{72}$ Mpc). 
The corresponding expected number of X-ray AGN 
($N_e$), was estimated by using the $\log N - \log S$ of Kim et al. (2007)
folding in the sensitivity map of each field.
The overdensity of X-ray sources in each cluster was then calculated
according to: $1+\delta_x = N_x/N_e$ and its uncertainty by using the small number 
Poisson approximation (eg. Gehrels 1986) for a confidence level corresponding to 
the 1$\sigma$ limit for Gaussian statistics (ie., 0.8413).
The robustness of the derived values of $\delta_x$ to uncertainties (a) of the $\log N - \log S$ relation and (b) 
of the overdensity determination method, was verified by varying the $\log N - \log S$ relation 
 within its quoted uncertainties and by 
alternatively using the overdensity estimation method (their eq.6) of Branchesi et al. (2007).
No significant variations of $\delta_x$ and of its uncertainty were found.

To calculate the corresponding overdensity of SDSS optical galaxies, 
and to minimize as much as possible projection effects,
we extracted all galaxies located within 5 $h^{-1}_{72}$ Mpc around
the cluster center (to facilitate the estimate also of the local 
background), having magnitudes in the range $m_r^*- 0.5 <m_r
<m_r^*+0.5$, where $m_r^*$ is the $r$-band magnitude corresponding to $M_r^*$ of the
Blanton et al. (2003) luminosity function at the redshift of the cluster. 
This characteristic magnitude, corresponding to the break of 
the luminosity function, is estimated by: $m^*= M^* + 5\log_{10} d_L + K(z) + 25 + A_\nu$,
with $d_L$ the luminosity distance of the cluster (for the ``concordance''
cosmology), $K(z)$ the K-correction for an elliptical galaxy SED
(based on Poggianti 1997)
and $A_\nu$ the Galactic absorption, calculated
using Galactic absorption maps of Schlegel et al. (1998).
We then counted the corresponding optical galaxies in the effective cluster
area and compared with a global and/or a local background estimate,
consistently for the corresponding magnitude range of each cluster. The 
former  was determined from a 20 sq.degrees region 
near the equatorial coordinate equator,
while the latter 
from the annulus between 4 to 5
$h^{-1}_{72}$ Mpc around each cluster center, a distance far enough 
to ensure negligible contamination by cluster galaxies. 
The overdensity results are robust to the different background
estimates, with the only exception of A2065, which appears to be embedded in an
overall high large-scale overdensity. We choose to present 
optical overdensity results based on the global background estimate.

Finally, the overdensity of optical galaxies in clusters was
calculated in the usual way, i.e., 
$1+\delta_o = N_o/\langle N\rangle$, with corresponding
Poisson uncertainty of:
$\sigma^2_{\delta_o}\simeq \Delta_o \left[1+\sigma_{b}^2 \Delta_o \right]$
where $\Delta_o=(1+\delta_o)^2/N_o$ and $\sigma_{b}$ is the Poisson uncertainty of the background 
optical galaxy density, with a typical value being $\sim 0.025$.


\begin{inlinefigure}
\centering
\psfig{file=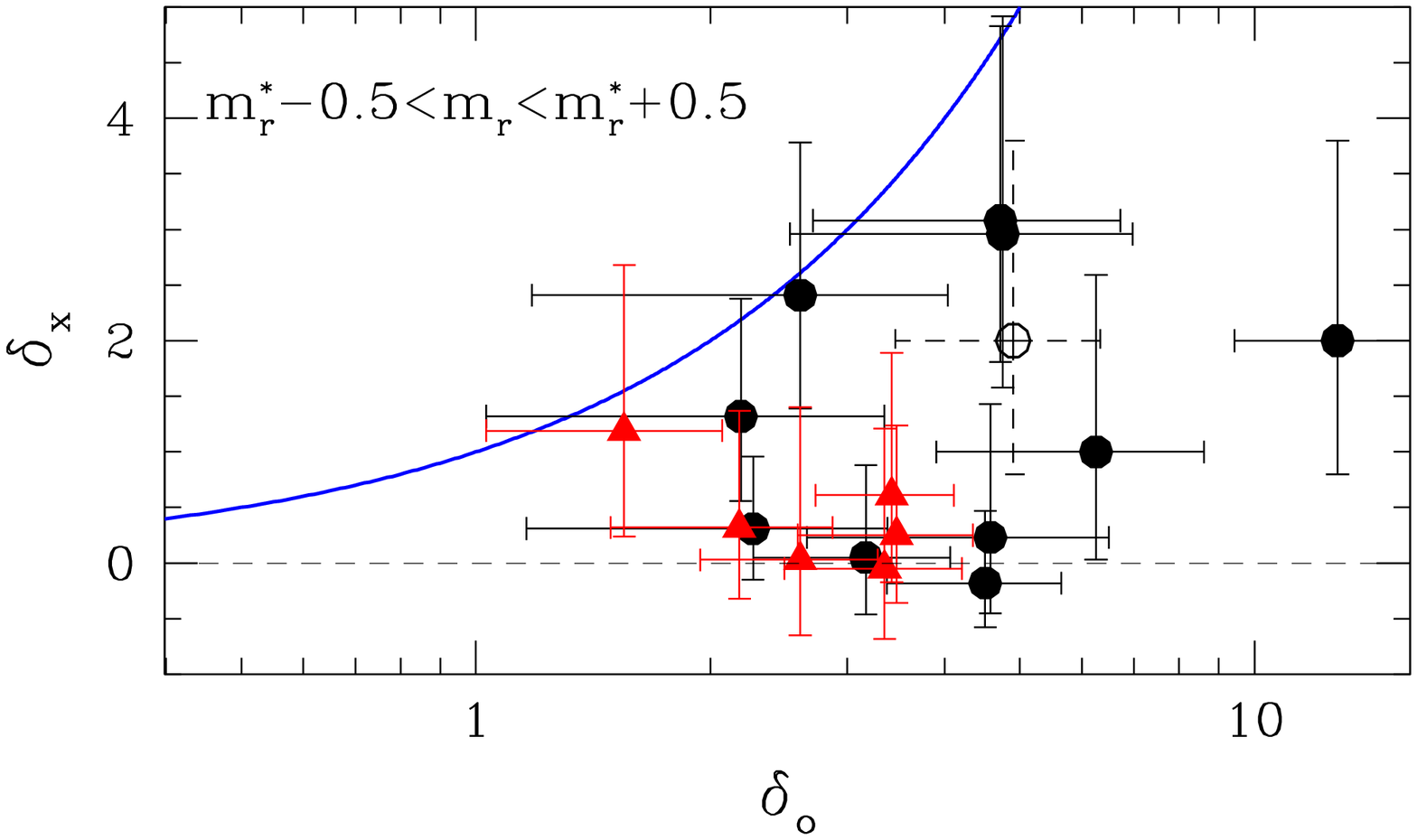,width=1.0\linewidth,clip=} 
\figcaption{X-ray AGN overdensity $\delta_x$ with respect to overdensity
of SDSS bright galaxies, $\delta_o$. Triangles denote the clusters
with flux limit corresponding to $L_x \magcir 2\times 10^{42}$erg
s$^{-1}$ (all $z\magcir 0.2$ clusters; see Table 1) 
while filled circles denote lower-$z$ clusters
with $L_x \simeq  10^{42}$erg $s^{-1}$. The empty dot corresponds to A2065 when using
its local background. The continuous curve 
corresponds to $\delta_x=\delta_o$, while the dashed line to $\delta_x=0$.
The errorbars correspond to Poisson uncertainties.}
\end{inlinefigure}

\section{Results \& Conclusions}
We find that there is a relatively significant X-ray source
overdensity in about half of the clusters in our sample, the rest showing 
the excepted background value of $\delta_x$. This can be realized by inspecting Table 1, where
we list the X-ray and optical overdensity values and their uncertainties, as well as from
 Figure 1 where we plot the cluster X-ray point-source overdensities versus the
corresponding optical SDSS galaxy overdensities, within the indicated $r$-band
magnitude range.
Note that the 6 clusters with effective X-ray
flux-limit corresponding to a
minimum luminosity of $L_x \magcir 2 \times 10^{42}$ erg s$^{-1}$ (see
specific values in Table 1), at the
redshift of the clusters, are indicated by 
a triangular point type.

Although the number of clusters in our sample is quite low to provide stringent
population statistics, a secure and important conclusion of our analysis is that 
the cluster X-ray point-source overdensities are always lower than
the corresponding optical SDSS overdensities (with only two clusters having 
$\delta_x\sim\delta_o$). 
In Table 2 we present such population statistics but separately for the two
cluster subsamples having different limiting $L_x$ (as discussed previously).
Note also that in order to take into account the variable uncertainty of the overdensity
values we present in Table 2 
the Poisson uncertainty-weighted mean optical and X-ray overdensities. 
A $t$-student test comparing the two means, assuming unknown and unequal
variances, shows that they are different at a high significance level (see ${\cal P}$ in Table 2).
Inspecting Table 2 it becomes evident that {\em the luminous X-ray AGN
overdensity is suppressed by a factor of 3 - 4, on the mean,
with what would have been expected from a constant fraction, 
independent of environment, of X-ray AGN to bright optical galaxies.}
This result is in the same direction with the
optical SDSS analysis of Lee et al. (2009). 
 
In order to investigate in more detail the nature of the X-ray
overdensities in our cluster sample, we have also cross-identified 
all detected X-ray sources
with the SDSS database, finding in total only six out of the 88
detected X-ray point-sources (with $L_x\ge 10^{42}$ erg s$^{-1}$ at
the redshift of the cluster) being clearly associated with the
clusters; among which 
one spectroscopically confirmed AGN (Sy1), 2 galaxies with no 
apparent emission lines 
and 3 more galaxies, based on their photometric redshifts (one of
which, in A1689, is indeed
confirmed by the spectroscopic analysis of Martini et al. 2007). 
In Table 1 we also present the optical characterization of the
X-ray point-sources, the total number
of which
for each cluster field, $N_x$, is listed in column 5. The 6$^{th}$ column
indicates the number of probable background QSO, $N_{\rm QSO}$, (mostly determined as such
from their point-like images and their 
$u-g$ versus $g-r$ colors, while $\sim 10\%$ are also
spectroscopically verified), the 7$^{th}$ column indicates
the number of X-ray sources clearly associated with cluster galaxies, $N_{\rm cgal}$, and the
8$^{th}$ column indicates the number of apparently irrelevant
associations, $N_{\rm other}$, like stars,
foreground/background galaxies, no-counterparts, smudges, etc.
About $\sim 50\%$ of all our AGN candidates 
appear to be related to
background QSO, which in many clusters represent the expected background provided 
by the $\log N-\log S$ of Kim et al. (2007).
Therefore, the real overdensity of X-ray AGN associated with our cluster sample 
appears to be
significantly smaller than what is listed in column 3 of Table 1, a fact which
further strengthens our main result that the rich cluster environment
(within 1 $h^{-1}_{72}$ Mpc) strongly suppresses the 
luminous X-ray AGN activity.

In order to provide also a visual example of the source
categorization, we present in Figure 2 the XMM field of A2065, 
the lowest $z$ cluster of our sample, 
together with the
SDSS composite $gri$-band images of all the X-ray point-source
counterparts.

Finally, we stress that the main conclusion of our present analysis is that
in the intermediate intracluster distances (ie., between $3r_c$ and 
1$h^{-1}_{72}$ Mpc), the relatively dense and hot ICM environment not 
only does not
enhance AGN activity, but it rather strongly suppresses it (at least
its X-ray luminous manifestation). 


{\small 
\acknowledgments
We would like to thank I.Georgantopoulos for discussions and the 
referee for useful suggestions.
M.Plionis acknowledges financial support under Mexican government CONACyT
grant 2005-49878.
Funding for the SDSS and SDSS-II has been provided by the Alfred P. Sloan
Foundation, the Participating Institutions, the National Science Foundation,
the U.S. Department of Energy, the National Aeronautics and Space
Administration, the Japanese Monbukagakusho, the Max Planck Society, and the
Higher Education Funding Council for England. The SDSS Web Site is
{\sc http://www.sdss.org/}. 
}

\begin{table}
\caption{Our cluster sample: X-ray and optical galaxy overdensities and  X-ray source categorization.}

\tabcolsep 6 pt
\begin{tabular}{lccccccccc} \\ \hline

Cluster & $z$  & $\delta_x$ & $L^{\rm limit}_x$/erg s$^{-1}$ 
& $N_x$ & $N_{\rm QSO}$ & $N_{\rm cgal}$ & $N_{\rm other}$ & $\delta_o$ & $N_g$$^{a}$ \\ \hline

A2065 & 0.073 &  2.00$^{+1.80}_{-1.20}$ &   1.0e42   & 6 & 4 & 1 & 1 &12.79$\pm$3.37$^{b}$ &109  \\
A1589 & 0.073 &  2.96$^{+1.96}_{-1.38}$ &   1.0e42   & 8 & 5 & - & 3 & 4.75$\pm$2.22 & 38 \\
A2670 & 0.076 &  1.00$^{+1.59}_{-0.97}$ &   1.0e42   & 4 & 1 & - & 3 & 6.26$\pm$2.36 &142 \\
A1663 & 0.084 &  3.08$^{+1.75}_{-1.27}$ &   1.0e42   & 10& 5 & - & 5 & 4.72$\pm$2.01 & 56  \\
A1750 & 0.085 &  1.23$^{+1.20}_{-0.68}$ &   1.0e42   & 3 & 2 & - & 1 & 4.58$\pm$1.92 & 40 \\
A1674 & 0.107 &  2.41$^{+1.37}_{-1.02}$ &   1.0e42   & 11& 5 & 1 & 5 & 2.61$\pm$1.43 &165 \\
A2050 & 0.118 &  1.32$^{+1.06}_{-0.76}$ &   1.0e42   & 9 & 2 & - & 7 & 2.19$\pm$1.17 & 50 \\
A1068 & 0.138 &  0.31$^{+0.65}_{-0.46}$ &   1.0e42   & 8 & 5 & 1 & 2 & 2.27$\pm$1.11 & 71 \\
A1689 & 0.183 & -0.18$^{+0.65}_{-0.40}$ &   1.0e42   & 4 & 2 & 1 & 1 & 4.51$\pm$1.14 &228 \\
A963  & 0.206 &  0.05$^{+0.83}_{-0.51}$ &   1.5e42   & 4 & 1 & - & 3 & 3.18$\pm$0.90 &134 \\
A773  & 0.217 &  0.25$^{+0.99}_{-0.61}$ &   1.8e42   & 4 & 2 & - & 2 & 3.47$\pm$0.88 &108 \\
A1763 & 0.223 & -0.05$^{+1.26}_{-0.63}$ &   2.5e42   & 2 & 1 & 1 & - & 3.35$\pm$0.86 &152  \\
A267  & 0.230 &  0.32$^{+1.05}_{-0.64}$ &   2.1e42   & 4 & 1 & - & 3 & 2.18$\pm$0.69 & 37  \\
A1835 & 0.253 &  0.03$^{+1.37}_{-0.68}$ &   3.5e42   & 2 & 1 & 1 & - & 2.61$\pm$0.67 & 48  \\
A2631 & 0.273 &  1.19$^{+1.49}_{-0.95}$ &   3.3e42   & 5 & 4 & - & 1 & 1.55$\pm$0.52 &136  \\
A1758 & 0.279 &  0.61$^{+1.28}_{-0.78}$ &   2.2e42   & 4 & 3 & - & 1 & 3.42$\pm$0.69 &198  \\ \hline

\tablenotetext{a}{$N_g$ is the Abell Richness, from Abell et al. (1989).}
\tablenotetext{b}{Using the local background estimate we obtain $\delta_o\simeq 4.9\pm 1.4$.}

\end{tabular}

\end{table}

\begin{table}
\caption{Population statistics for our cluster sample.}

\tabcolsep 6 pt
\begin{tabular}{lccccc} \\ \hline

$L^{\rm limit}_x$ (erg s$^{-1}$) & $\#$ & $\langle \delta_x/\delta_o\rangle$  & 
$\langle \delta_x\rangle_w$  & $\langle \delta_o\rangle_w$  & $1-{\cal P}$\\ \hline
$1\times 10^{42}$ & 10 & $0.33\pm 0.34$ 
& $0.91\pm0.76$ & $3.99\pm 1.89$ & 0.995\\
$\magcir 2\times 10^{42}$ & 6 & $0.19\pm 0.29$ 
& $0.36\pm0.34$ & $2.65\pm 1.10$ & 0.999\\ \hline

\tablenotetext{1}{$\langle \delta_x\rangle_w$ and $\langle \delta_o\rangle_w$ 
correspond to Poisson uncertainty-weighted means.}
\tablenotetext{2}{${\cal P}$ is the probability that the unweighted 
$\langle \delta_o \rangle$ and $\langle \delta_x \rangle$ are equal. The 
corresponding probability for the weighted means is even larger.}
\end{tabular}

\end{table}

\clearpage
\begin{inlinefigure}
\centering
\psfig{file=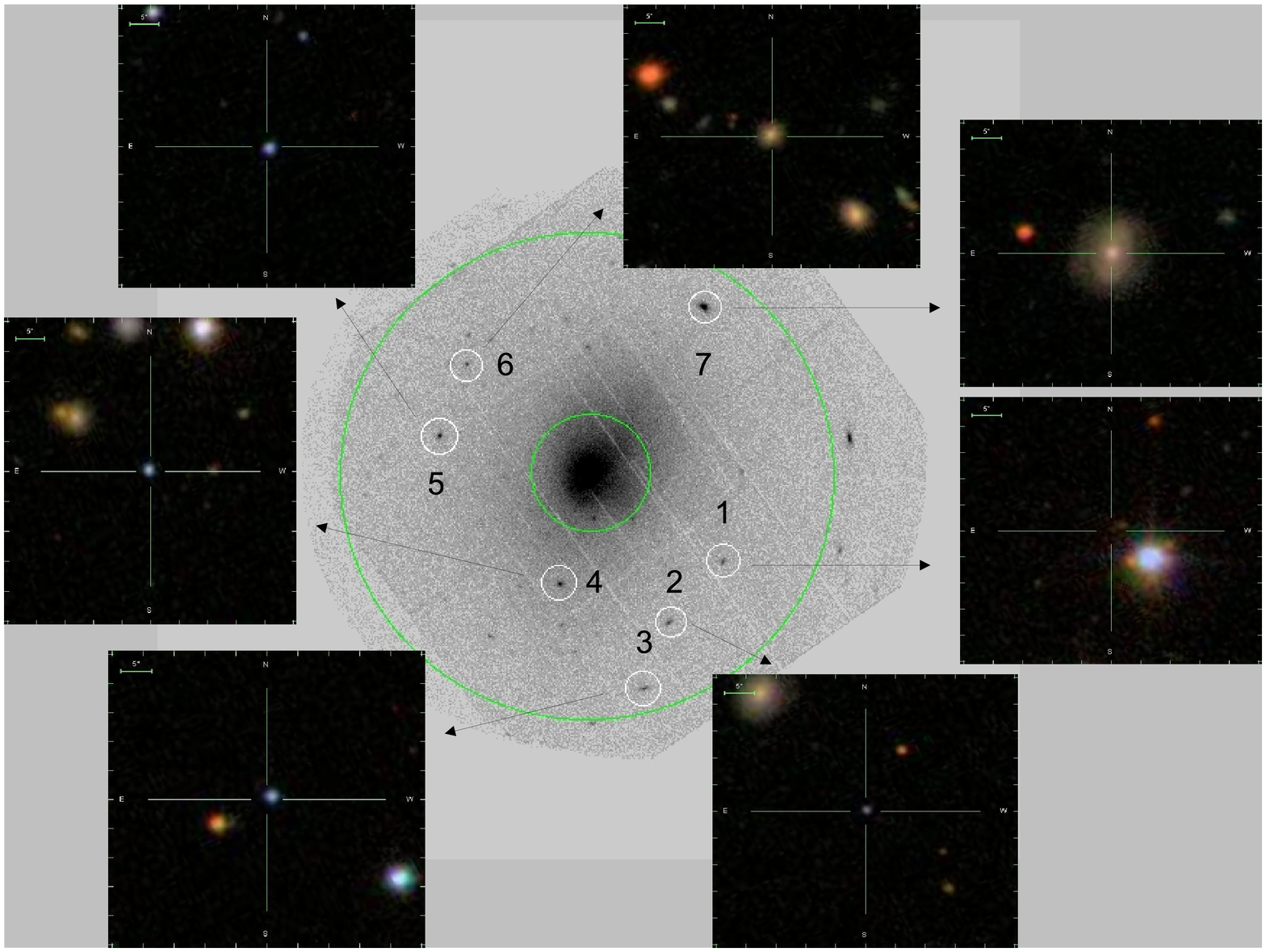,width=0.9\linewidth,clip=} 
\figcaption{An example of the X-ray image of a cluster field from our
  sample (Abell 2065). 
The green circles delineate the inner and outer radii, within which we measure
overdensities. The SDSS $gri$-composite image possibly corresponds to a 
star. Images 2, 3, 4 and 5 are probable background QSO, image 6
corresponds to a cluster galaxy (as indicated by its photometric
redshift) but with $L_x<10^{42}$ erg s$^{-1}$ (and thus not included in our analysis), while image 7
  corresponds to a $L_x\ge 10^{42}$ erg s$^{-1}$ Seyfert 1 belonging to the cluster ($z=0.0747$).
}

\end{inlinefigure}

\end{document}